\documentclass[onecollarge,sort&compress,natbib]{svjour2}
\usepackage{braket}
\usepackage{dsfont}
\usepackage{amsmath}
\usepackage{amssymb}
\usepackage{mathtools}
\usepackage{graphicx}
\usepackage{dcolumn}
\usepackage{bm}
\usepackage{multirow}
\usepackage{leftidx}
\usepackage{color}

\usepackage[german,english]{babel}
\makeatletter
\newcommand{\vast}{\bBigg@{4}}
\newcommand{\Vast}{\bBigg@{5}}
\makeatother
\bibpunct{[}{]}{;}{n}{}{,} 
\smartqed  
\journalname{Few-Body Systems}

\begin{document}
\title{Van der Waals universality in homonuclear atom-dimer elastic collisions}

 \author{P. Giannakeas \and Chris H. Greene}

 
 \institute{P. Giannakeas \at
              Department of Physics and Astronomy, Purdue University, West Lafayette, Indiana 47907, USA \\
              \email{pgiannak@purdue.edu}           
           \and
           Chris H. Greene \at
             Department of Physics and Astronomy, Purdue University, West Lafayette, Indiana 47907, USA \\
              \email{chgreene@purdue.edu} 
}

 \date{\today}

\maketitle
\begin{abstract}
The universal aspects of atom-dimer elastic collisions are investigated within the framework of Faddeev equations.
The two-body interactions between the neutral atoms are approximated by the separable potential approach.
Our analysis considers a pure van der Waals potential tail as well as soft-core van der Waals interactions permitting us in this manner to address the universally general features of atom-dimer resonant spectra.
In particular, we show that the atom-dimer resonances are solely associated with the {\it excited} Efimov states.
Furthermore, the positions of the corresponding resonances for a soft-core potentials with more than 5 bound states are in good agreement with the corresponding results from an infinitely deep pure van der Waals tail potential.
\end{abstract}


\section{Introduction}

Vitaly Efimov in the 1970s conveyed the concept of three-body bosonic bound states where the two-body subsystems are unbound \cite{eflmov1971weakl}.
Furthermore, within a zero-range model Efimov showed that at unitarity there is an infinity of such trimer states whereas the corresponding binding energies scale exponentially.
Despite the appealing simplicity of the concept of Efimov states from a theoretical viewpoint, their experimental observation only occurred about four decades later through advances in the realm of ultracold atomic physics \cite{kraemer2006evidence}.
This experimental realization sparked extensive theoretical \cite{fedorovprl1993,esry1999recombination,bedaque2000three,suno2002three,suno2003recombination,petrovprl2004,Braaten2007120,braaten2006universality,Hammer2007,von2009signatures,incaophysjb2009,wangprl2012,wang2011efimov,naidonpra2014,naidonprl2014,blume2015efimov} and experimental \cite{zaccanti2009observation,williamsprl2009,knooppra2012,Ferlainoprl2009,bernigerprl2011,Lompe940,Nakajimaprl2011,lompeprl2010,Nakajimaprl2010,knoop2009observation, Pollack1683,grossprl2009,wenzpra2009,barotiniprl2009,huckansprl2009,zenesininjp2013,grossprl2010,ferlainofewbody2011,zenesinipra2014,kunitski2015observation,machteyprl2012} interest in order to develop a comprehensive understanding of the underlying physics of Efimov states.
One intriguing aspect of the Efimov physics is that within the zero-range approximation the corresponding spectrum is not bounded from below, which is sometimes referred to as the ``Thomas collapse'' which is associated with the zero-range model Hamiltonian.
By introducing an additional length scale, i.e. the three body parameter, the Efimov spectrum possesses a ground state with finite energy.
Various theoretical studies \cite{eflmov1971weakl,Braaten2007120,braaten2006universality} pointed out that the Efimov spectrum is inherently system dependent due to the three body parameter.
However, experimental evidences \cite{ferlainofewbody2011} on three-body collisions between ultracold atoms manifest that the resulted three-body spectrum possesses universal characteristics associated with the van der Waals interactions which the neutral atoms experience.
The microscopic mechanism associated with the universality of the three-body parameter in ultracold collisions was shown theoretically to be related with the sudden drop of the two-body van der Waals interaction potential as the interparticle distance decreases, which in return produces an effective repulsive three-body potential that prevents the three neutral atoms from approaching to hyperradial distances $R<2\ell_{\rm vdW}$ ($\ell_{\rm vdW}$ is the van der Waals length scale) \cite{wangprl2012,naidonpra2014,naidonprl2014}.

In view of the theoretical efforts to understand the underlying universal mechanisms of three-body bound systems it is of interest to extent the concept of van der Waals universality to atom-dimer collisional systems where a boson elastically collides on a pair of bound particles.
In particular, one main question that immediately arises is whether an atom-dimer system possesses resonant spectra that exhibit universal characteristics.
In order to address this question in this work the elastic atom-dimer collisions are investigated beyond the zero-range approach \cite{Braaten2007120,braaten2006universality,helfrich2009resonant,kievsky2014universality,hammer2010efimov}.
In the spirit of a simple model, the theoretical framework of Faddeev equations is employed in our study and the two body interactions are treated within the separable potential approximation.
In particular two types of interactions are considered, namely a van der Waals tail potential as well as soft-core van der Waals two body interactions.
The case of the pure van der Waals tail potential is modeled by the separable potential approximation introduced by Naidon {\it et al.} \cite{naidonpra2014} whereas the soft-core van der Waals interactions are factorized by employing properties of the Hilbert-Schmidt approach  \cite{kharchenko1969three, sitenko2013lectures,schmid2013quantum}.
Note that for the case of the universal three body bound states the Naidon {\it et al.} separable potential approximation was shown to encompass the key aspects of the van der Waals physics that constrain the 3-body parameter.
Therefore, these types of potentials allow us to systematically explore the universal aspects of the corresponding atom-dimer resonant spectra.
More specifically, our calculations show that in the regime of low energies the atom-dimer resonances occur only for the excited Efimov states and not for the ground Efimov state.
In addition, we observe that for deep soft-core potential containing more than 5 bound states the corresponding atom-dimer resonances are in good agreement with the resonances for the Naidon {\it et al.}  separable potential approximation which corresponds to an infinitely deep van der Waals potential.
Namely, in this case the first atom-dimer resonance occurs at scattering lengths $a_s\sim5.618\pm0.013 \ell_{vdw}$ whereas the second one occurs at $a_s\sim192.5\pm 0.3 \ell_{vdw}$.

\section{Low energy atom-dimer collisions}
Consider a system of three identical neutral bosonic atoms which collide at low energies.
More specifically, it is assumed that a bosonic pair forms a bound state with dimer energy $E_d$, whereas the third atom has just barely enough kinetic energy to collide with the dimer.
The two-body interactions between the neutral atoms are described by a van der Waals potential $V_{\rm{vdW}}(\boldsymbol{r})$.
Due to low collisional energies only the $s$-wave interactions are assumed to be important between each pair of atoms.
In addition, the total colliding energy of the atom-dimer system is considered to be below the three-body breakup threshold and of course just above the dimer threshold, namely $E_d\leq E<0$.

In order to explore the universal aspects of the atom-dimer collisions the Faddeev equations are employed.
In particular, the framework of the Faddeev equations permits us to derive the full off-shell scattering amplitudes for the atom-dimer system \cite{schmid2013quantum,sitenko2013lectures,watson1967topics,braaten2001three,bedaque2003low,bedaque1999renormalization,bedaque2002review,bedaque1999three}.
Moreover, in the spirit of a simple model our analysis is restricted to the separable potential approximation \cite{yamaguchiphysrev1954}, in which the two-body interaction for $s$-wave collisions is expressed as follows:
\begin{equation}
 \hat{V}_{\rm{vdW}}=-\frac{\hbar^2 \lambda}{m}\ket{\chi}\bra{\chi}
 \label{eq1}
\end{equation}
where $\lambda$ denotes the strength of the two body potential, $m$ indicates the mass of the atom and $\ket{\chi}$ represents the form factor.
Note that the explicit expression of the form factor $\ket{\chi}$ is constrained by the nature of the low energy two-body physics.
In particular, for $\braket{\boldsymbol{q}|\chi}\to 1$, where $\boldsymbol{q}$ refers to the relative momentum between two particles, the corresponding potential describes the collisional aspects of two particles interacting with zero-range interactions.
Note that in this approximation the corresponding three-body Faddeev equations reduce to the Skorniakov-Ter-Martirosian equation \cite{skorniakovjetp1957}.
In order to include the two-body effective range corrections a different expression for the form factors must be adopted \cite{shepard2007calculations}.
In our study, the focus is on neutral atoms which interact via a van der Waals potential, hence the form factor $\ket{\chi}$ is constructed such that the long range behavior of the $V_{\rm{vdW}}(\boldsymbol{r})$ is included.
A more detailed discussion about this aspect is presented in subsection \ref{sect:spvdw}.
An important attribute of the separable potential approximation is that the corresponding Faddeev equations are significantly simplified yielding one dimensional integral equations that can be solved efficiently and accurately.

For completeness, the following subsection briefly reviews the three body off-shell scattering amplitude integral equations for elastic atom-dimer collisions, adopting the notation based mainly on Ref.\cite{afnan2004three}.

\subsection{ Faddeev equations for elastic atom-dimer collisions}
In order to tackle the elastic atom-dimer collisions, the three-body scattering amplitudes are expressed in momentum space.
In particular, the relative motion between dimer bosons is described by the $\boldsymbol{q}$ momentum whereas the $\boldsymbol{p}$ is the momentum of the spectator boson with respect to the pair.
Then the integral equation for the three-body scattering amplitude reads:
\begin{equation}
 X(\boldsymbol{p},\boldsymbol{p}';E^+)=2 Z(\boldsymbol{p},\boldsymbol{p}';E^+)+2 \int \frac{d \boldsymbol{p}''}{(2 \pi)^3}Z(\boldsymbol{p},\boldsymbol{p}'';E^+) \tau(E^+-\frac{p''^2}{2 M})X(\boldsymbol{p}'',\boldsymbol{p}';E^+),
\label{eq2}
\end{equation}
where $M=2m/3$ is the reduced mass of the spectator boson with respect to the dimer and $m$ is the mass of the atom.
$\tau(\cdot)$ indicates the two-body transition matrix embedded in the three-body momentum space.
$E^+=E+i \epsilon$ denotes the total colliding energy including an positive infinitesimal quantity $\epsilon$.
Note that the infinitesimal $\epsilon$ enforces outgoing wave boundary conditions and permits closed-form contour integration of the poles which are contained in the amplitude $Z(\cdot,\cdot;\cdot)$ and the dressed atom-dimer propagator $\tau(\cdot)$.
Note that the amplitude $X(\cdot,\cdot;\cdot)$ describes the elastic collision of an atom with a dimer.

The atom-dimer scattering amplitude in Eq.~(\ref{eq2}) can be expanded in partial-waves for total angular momentum $L=0$ \cite{schmid2013quantum,sitenko2013lectures,watson1967topics}. 
Since all the three atoms interact with only $s$-wave interactions the corresponding form of the atom-dimer scattering amplitude is given by the following expression:
\begin{equation}
 X_0(p,p';E^+)=2 Z_0(p,p';E^+)+2 \int_0^\infty \frac{d p''}{2\pi^2} p''^2Z_0(p,p'';E^+) \tau_0(E^+-\frac{p''^2}{2 M})X_0(p'',p';E^+).
\label{eq3}
\end{equation}

After implementing the separable potential given in Eq.~(\ref{eq1}), the first Born amplitude $Z_0(\cdot,\cdot;\cdot)$ in Eq.~(\ref{eq3}) is given by the following expression:
\begin{equation}
  Z_0(p,p';E^+) = \frac{1}{2}\int^{1}_{-1}d\xi\frac{\chi(|\boldsymbol{p}+\frac{\boldsymbol{p}'}{2}|)\chi(|\boldsymbol{p}'+\frac{\boldsymbol{p}}{2}|)}{E^+-\frac{p^2}{m}-\frac{p'^2}{m}-\frac{\boldsymbol{p}\cdot\boldsymbol{p}'}{m}},
\label{eq4}
\end{equation}
where $\xi=\hat{\boldsymbol{p}}\cdot\hat{\boldsymbol{p}}'$ is the angle between the momenta $\boldsymbol{p}$ and $\boldsymbol{p}'$.

As was mentioned at the beginning of the section, the total colliding energy is always negative in this study, and below the breakup threshold. 
This implies that the $Z_0$ Born amplitudes are free of poles; however, the energy is above the dimer threshold meaning that the dressed two-body propagator $\tau_0$ (see Eq.~(\ref{eq3})) has poles which should be integrated out.
In order to show this, the two-body transition matrix for a separable potential is demonstrated whereas its pole structure is explicitly isolated. 
The full off-shell two-body transition matrix with separable interactions reads:
\begin{equation}
 \braket{\boldsymbol{q} |\hat{t} |\boldsymbol{q}'}=\chi(\boldsymbol{q}) \tau_0(E) \chi(\boldsymbol{q}'),
\label{eq5}
\end{equation}
where the dressed propagator $\tau_0(E)$ is given by the following expression:
\begin{equation}
 \tau_0(E)=\frac{S(E)}{E-E_d}~~{\rm{with}}~S^{-1}(E)=\int_0^\infty \frac{d q}{2 \pi^2} q^2\frac{\chi(q)^2}{(E-\frac{q^2}{m})(E_d-\frac{q^2}{m})},
 \label{eq6}
\end{equation}
where at $E=E_d$ the dressed propagator diverges, while the quantity $S(E)$ for $E\to E_d$ is free of poles since $E_d\leq0$.

Since the pole structure of the dressed propagator has been isolated, Eq.~(\ref{eq6}) is substituted in Eq.~(\ref{eq3}), which produces the following expression for the atom-dimer scattering amplitude:
\begin{equation}
 X_0(p,p';E^+)=2 Z_0(p,p';E^+)+2 \int_0^\infty \frac{d p''}{2\pi^2} p''^2Z_0(p,p'';E^+)\frac{S(E^+-\frac{3p''^2}{4 m})}{E^+-\frac{3p''^2}{4 m}-E_d}	X_0(p'',p';E^+).
\label{eq7}
\end{equation}

By inspection of Eq.~(\ref{eq7}) it is evident that the second term always diverges.
Specifically, when the spectator particle is far from the dimer either before or after the collision the total energy is distributed into the binding energy of the dimer and the kinetic energy of the spectator atom, namely $E^+=E_d+\frac{3 k^2}{4 m}+i 0$, where $k>0$.
Because of the latter the denominator is proportional to $\sim(k^2-q^2+i 0)$ yielding a singular integrand in Eq.~(\ref{eq7}).
One way to treat the pole of equation is to convert the integral equation in Eq.~(\ref{eq7}) into a principal value integral equation through the identity $ \frac{1}{E^+-\frac{3p''^2}{4 m}-E_d}=\mathcal{P}\frac{1}{E-\frac{3p''^2}{4 m}-E_d}-i\pi \delta(E-\frac{3p''^2}{4 m}-E_d)$. 
Then the atom-dimer scattering amplitude in Eq.~(\ref{eq7}) is converted into an off-shell K-matrix for elastic atom-dimer collisions.
Note that the above mentioned $K$-matrix is not the {\it conventional} reaction or K-matrix of formal scattering theory. 
Indeed, as we show below the K-matrix of Eq.~(\ref{eq8}) is proportional to the tangent of the atom-dimer phase shift, i.e. $\tan \delta_{AD}$.
The principal value integral equation for this atom-dimer $K$-matrix is given by the following expression:
\begin{equation}
 K(p,p';E)=2 Z_0(p,p';E)+2 \mathcal{P}\int_0^\infty \frac{d p''}{2\pi^2} p''^2Z_0(p,p'';E)\frac{S(E-\frac{3p''^2}{4 m})}{E-\frac{3p''^2}{4 m}-E_d}K(p'',p';E),
\label{eq8}
\end{equation}
where the off-shell $K$-matrix is real, symmetric and free of poles. These attributes make its numerical implementation and solution rather straightforward.

As stated above the atom-dimer $K$-matrix is the off-shell one, however only its on-shell part is of physical significance and can be associated with the corresponding atom-dimer scattering length.
The notion of the on-shell $K$-matrix basically means that the Jacobi momenta $p$ and $p'$ are equal to the relative momentum $k$ between the spectator atom and the interacting pair, namely $p=p'=k$.
Recall, that the total colliding energy of the atom-dimer system is $E=E_d+\frac{3 k^2}{4 m}$ and that we are interested in the elastic scattering of an atom from a pair of particles.
Assuming that the $s$-wave atom-dimer phase shift is $\delta_{\rm{AD}}$, then the on-shell $K$-matrix obeys the following relation:
\begin{equation}
 K(k,k;E)=-\frac{3}{2 m \pi} \frac{\tan \delta_{\rm AD}}{k}  S^{-1}(E_d).
 \label{eq9}
\end{equation}
Here the quantity $-k^{-1} \tan \delta_{\rm{AD}}$ defines a {\it generalized energy-dependent} atom-dimer scattering length, $a_{\rm{AD}}(k)$, for total collision energy $E=E_d+\frac{3 k^2}{4 m}$.
Using Eq.~(\ref{eq9}) the atom-dimer $s$-wave scattering length $a_{\rm{AD}}(k)$ can be expressed in terms of the on-shell $K$-matrix.
\begin{equation}
  a_{\rm{AD}}(k)=\frac{2 m \pi}{3} S(E_d)K(k,k;E).
\end{equation}

Having briefly reviewed the off-shell $K$-matrix principal value integral equations for elastic atom-dimer collisions, and having introduced the energy dependent $s$-wave atom-dimer scattering length in terms of the on-shell $K$-matrix, the following subsection focuses on the construction of the separable potential which fully encapsulates the two-body s-wave van der Waals physics of the neutral atoms.

\subsection{Separable potentials for two-body van der Waals interactions} \label{sect:spvdw}

The separable potential approximation plays a key role in the simplification of the Faddeev equations.
However, it should be noted that a separable potential is inherently non-local in contrast to the {\it true} two-body interactions which are local.
Difficulties associated with the nonlocality can largely be circumvented by designing the $\ket{\chi}$ form factors of a separable potential such that they describe as much of the two-body physics of two interacting neutral atoms as is manageable.
Indeed, Naidon {\it et al} in Refs.\cite{naidonpra2014,naidonprl2014} highlight the role of the separable potential approximation designed specifically to include the two-body van der Waals physics.
These two studies provide further physical insight on the universal aspects of the spectrum of Efimov states \cite{ferlainofewbody2011} for three neutral atoms with equal masses.

Therefore, in order to study the universal characteristics of elastic atom-dimer collisions, the following subsection reviews the separable potential of Naidon {\it et al.} and adopts an alternative method to construct van der Waals separable potentials based on the Hilbert-Schmidt expansion \cite{kharchenko1969three, sitenko2013lectures,schmid2013quantum}.
The latter technique is applied in the case of a soft-core van der Waals potential, which enables a straightforward comparison with the Naidon {\it et al.} approach. Using this implementation, we study in detail how the atom-dimer scattering length depends on the {\it total} number of bound states which the soft-core two-body interactions support.
Note that both types of separable potentials contain information only for the least bound state which is supported from the {\it real} van der Waals interactions.
This constraint arises from the fact that the corresponding atom-dimer system experiences only elastic collisions, since inelastic processes associated with deeper two-body bound states are not included in the model.
Consequently, both separable potentials have the same functional form as is given in Eq.~(\ref{eq1}) 

\subsubsection{The Naidon {\it et al.} separable potential for van der Waals interactions}

Naidon {\it et al} in Refs.\cite{naidonpra2014,naidonprl2014} consider only a van der Waals potential tail, namely
$V_{\rm vdW}(\boldsymbol{r})=-C_6/r^6$, where $C_6$ is the dispersion coefficient and $r$ refers to the relative distance between two neutral atoms.
The corresponding separable potential is constructed by employing the {\it zero-energy} two-body wavefunction which is known analytically \cite{flambaumpra1999} and obeys the following relation:
 \begin{equation}
 \phi(r)= \Gamma\left(\frac{5}{4}\right)\sqrt{\frac{r}{\ell_{\rm vdW}}} J_{\frac{1}{4}}\left(2 \frac{\ell_{\rm vdW}^2}{r^2}\right)-\frac{\ell_{\rm vdW}}{a_s}\Gamma\left(\frac{3}{4}\right)\sqrt{\frac{r}{\ell_{\rm vdW}}} J_{-\frac{1}{4}}\left(2 \frac{\ell_{\rm vdW}^2}{r^2}\right),
 \label{eq11}
 \end{equation}
 where $ \ell_{\rm vdW}=\frac{1}{2}(m C_6/\hbar^2)^{1/4}$ indicates the van der Waals length scale and $a_s$ is the $s$-wave scattering length.
 The quantities $\Gamma(\cdot)$ and $J_{\pm \frac{1}{4}}(\cdot)$ represent the Gamma and Bessel functions, respectively.
Note that this wavefunction at large distances behaves as $\phi(r)\to 1-r/a_s$, whereas at short length scales it possesses fast oscillations and vanishes at the origin.

In momentum space a separable potential of the form given in Eq.~(\ref{eq1}) obeys the relation
\begin{equation}
 V_{\rm NP}(\boldsymbol{q},\boldsymbol{q}')= - \frac{\hbar^2 \lambda}{m}\chi(\boldsymbol{q})\chi(\boldsymbol{q}'),
\label{eq12}
\end{equation}
where the abbreviation ``NP`` stands for the Naidon {\it et al.} potential, $\lambda$ indicates the strength of the potential, $m$ denotes the mass of the atom, and the form factor $\chi(\boldsymbol{q})$ is expressed in terms of the zero-energy two-body wavefunction (see Eq.~(\ref{eq11}))
\begin{equation}
 \chi(\boldsymbol{q})\equiv\chi(q) =1-q\int_0^\infty d r \left[1-\frac{r}{a_s}-\phi(r)\right]\sin(q r).
 \label{eq13}
\end{equation}
Due to the $s$-wave character of the two-body interactions the form factor depends only on the magnitude of the momentum.
The strength of $V_{\rm NP}$ is also self consistently expressed in terms of the zero-energy two-body wavefunction.
\begin{equation}
 \lambda=\left[-\frac{1}{4\pi a_s}+\frac{1}{2\pi^2}\int_0^\infty d q |\chi(q)|^2 \right]^{-1}.
 \label{eq14}
 \end{equation}

 \subsubsection{Separable potentials for soft-core van der Waals interactions}
Consider that two neutral atoms interact with the following soft-core potential:
\begin{equation}
 V_{\rm SC}(r)=-\frac{C_6}{r^6+\sigma^6},
 \label{eq15}
 \end{equation}
 where the abbreviation ''SC`` stands for ''soft-core`` $\sigma$ indicates a quantity that controls the depth of the potential and {\it regularizes} it for $r\to0$.

 Moreover, assuming only $s$ partial wave interactions the soft-core potential in momentum space obtains the form
 \begin{equation}
  V^{s}_{{\rm SC}} (q,q')=4\pi \int_0^\infty d r~ r^2 j_0(q r) V(r) j_0(q' r),
  \label{eq16}
  \end{equation}
  where $j_0(\cdot)$ denotes the $s$-wave spherical Bessel function, and superscript $s$ indicates that the $s$-wave orbital angular momentum of the two interacting atoms, namely $\ell=0$.

The local potential of Eq.~(\ref{eq16}) is desirable to be expressed in terms of separable potentials, i.e. factorized in momentum space.
As was mentioned previously the main aim is to construct a separable potential for the soft-core van der Waals interactions which contains {\it all} the relevant information of the {\it least} bound state supported by the potential of Eq.~(\ref{eq15}).
This is achieved here by exploiting the Hilbert-Schmidt approach and its properties.
More specifically, our goal is to obtain the eigenfunctions and eigenvalues of the non-symmetric two-body Lippmann-Schwinger integral equation which reads:
  \begin{equation}
    \hat{V}^{s}_{{\rm SC}} \hat{G}_0(E) \ket{g_\nu;E}=\eta_\nu(E) \ket{g_\nu;E},
\label{eq17} 
\end{equation}
where $\hat{G}_0(E)=[E-H_0]^{-1}$ is the two-body Green's function and $H_0$ indicates the kinetic energy operator.
The quantity $E$ is assumed to be negative, because we are interested in the case where the two particles can form a bound state. 
Recall that the main purpose is to derive a separable van der Waals potential which then is used to describe elastic atom-dimer collisions.
Finally, $\eta_\nu(E)$ refers to the $\nu$-th eigenvalue of the non-symmetric Lippmann-Schwinger kernel, where $\nu$ is an integer labeling the different eigenvalues in a descending order.

The orthonormalization condition for the eigenvectors $\ket{g_\nu;E}$ is given by the following relation:
\begin{equation}
 \braket{g_{\nu'};E|G_0(E)|g_{\nu};E}=-\delta_{\nu'\nu}.
 \label{eq18}
 \end{equation}

One main property of the eigenvalues of Eq.~(\ref{eq17}) is that they should fulfill the condition $\eta_\nu(E)=1$ when $E=E_n$, where $E_n$ corresponds to the energy of the {\it least } bound state of a soft-core van der Waals potential with $n$ bound states in total.
Therefore, by fixing the parameter $\sigma$ of the potential in Eq.~(\ref{eq16}) the total number of bound states is defined.
Then, in the momentum representation the integral equation in Eq.~(\ref{eq17}) is diagonalized by varying the energy $E$ in order to obtain the eigenvalue $\eta_\nu$ which obey the condition $\eta_\nu(E=E_n)=1$. 
The corresponding eigenvector $\ket{g_\nu;E_n}$ is the form factor associated with the $n$-th {\it least} bound state of $V_{\rm SC}$.
The two-body wavefunction of the shallowest bound state and the eigenvectors $\ket{g_\nu;E_n}$ are connected in the momentum representation according to the following relation:
\begin{equation}
 \phi_{n,\nu}(q)=N_n\frac{g_\nu(q;E_n)}{\frac{\hbar^2 q^2}{m}-E_n},~~{\rm with}~E_n<0
 \label{eq19}
\end{equation}
where $N_n$ is the normalization constant of the $n$-th bound state.

The separable potential associated with the wavefunction in Eq.~(\ref{eq19}) then has the following form:
\begin{equation}
 \bar{V}_{SC}(q,q')=- g_\nu(q;E_n)g_\nu(q’;E_n),
 \label{eq20}
\end{equation}
where it should be noted that the form factors in the preceding equation are proportional to two-body wavefunction at energy $E_n$ whereas in the Naidon {\it et al.} separable potential the corresponding form factor is proportional to the zero-energy two-body wavefunction only. 

\begin{figure}[t!]
\centering
 \includegraphics[scale=0.65]{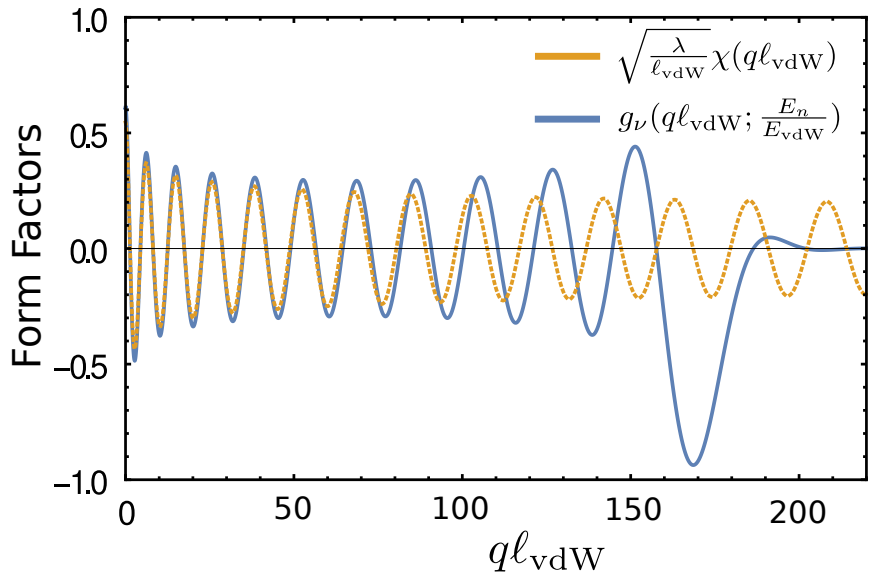}
 \caption{(color online) An illustration of the scaled form factor $\sqrt{\frac{\lambda}{ \ell_{\rm vdW}}}\chi(q\ell_{\rm vdW})$ given in Eqs.~(\ref{eq13}) and (\ref{eq14}) (orange dashed line) in the Naidon {\it et al.} separable approximation. $g_\nu(q \ell_{\rm vdW};\frac{E_n}{E_{\rm vdW}})$ (blue solid line) indicates the form factor for the soft-core vdW potential which supports 22 bound states. In both cases the scattering length is $a_s=32.3873 \ell_{\rm vdW}$.
 Note that $\ell_{\rm vdW}=\frac{1}{2}(m C_6/\hbar^2)^{1/4}$ is the van der Waals length scale and $E_{\rm vdW}=\frac{\hbar^2}{m\ell_{\rm vdW}^2}$ refers to the van der Waals energy scale.
 }
\label{fig0}
\end{figure}

Fig.\ref{fig0} compares the form factors obtained for the separable potentials $V_{\rm NP}(q,q')$ and $\bar{V}_{SC}(q,q')$ with the $s$-wave scattering length set to $a_s=32.3873 \ell_{\rm vdW}$.
In particular, the orange dashed line denotes the scaled form factor  $\sqrt{\frac{\lambda}{ \ell_{\rm vdW}}}\chi(q\ell_{\rm vdW})$ [see Eq.~(\ref{eq13})] for the Naidon {\it et al.} separable potential.
The blue solid line indicates the form factor  $g_\nu(q \ell_{\rm vdW};\frac{E_n}{E_{\rm vdW}})$ which corresponds to the least bound state of a soft-core potential with 22 bound states in total.
At large momenta $q$ the $g_\nu(q \ell_{\rm vdW};\frac{E_n}{E_{\rm vdW}})$ vanishes whereas  $\sqrt{\frac{\lambda}{ \ell_{\rm vdW}}}\chi(q\ell_{\rm vdW})$ possesses an oscillatory behavior with vanishing amplitude.
This behavior of $\sqrt{\frac{\lambda}{ \ell_{\rm vdW}}}\chi(q\ell_{\rm vdW})$ mainly arises from the fact that the form factors in Naidon {\it et al.} separable potential correspond to the least bound state of an infinitely deep vdW potential tail.
In addition, at small momenta $q$ we observe that both form factors, i.e.  $\sqrt{\frac{\lambda}{  \ell_{\rm vdW}}}\chi(q\ell_{\rm vdW})$ and $g_\nu(q \ell_{\rm vdW};\frac{E_n}{E_{\rm vdW}})$, possess the same nodal structure.
This is indicative of the van der Waals universality, since the small momenta behavior in spatial space refers to large separation distances which is strongly characterized by the attractive van der Waals potential tail.

Summarizing, in this subsection two types of separable potentials for vdW interactions are constructed, one is based on a hard wall, i.e. Naidon's separable potential [(see Eq.~\ref{eq12})] and the other one is based on a soft-core vdW potential [See Eq.~(\ref{eq20})].
In the following, these two types of separable potentials will be employed to calculate atom-dimer elastic collisions in order to extract their universal aspects.

\section{results and discussion}

\begin{figure}[t]
\centering
 \includegraphics[scale=0.65]{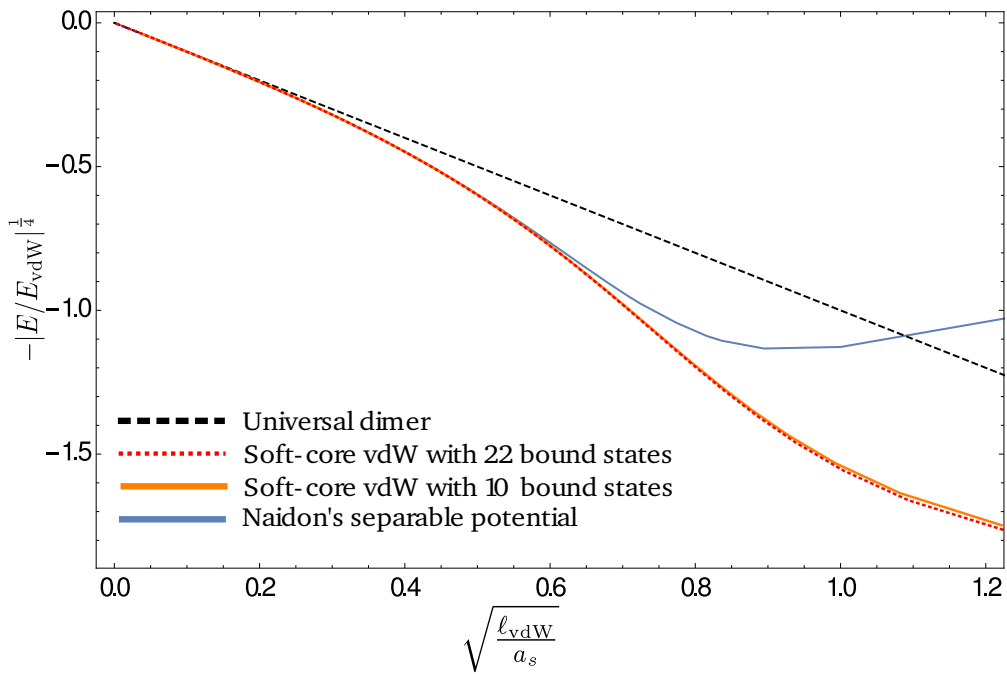}
 \caption{(color online) The dimer binding energies as a function of the two-body $s$-wave scattering length. The black dashed line indicates the universal dimer energies. The blue solid line corresponds to the dimer energies based on the Naidon {\it et al.} separable potential. The red dotted and orange solid line refer to the calculations for soft-core vdW potential which contain in total 22 and 10 bound states, respectively. Note that $\ell_{\rm vdW}=\frac{1}{2}(m C_6/\hbar^2)^{1/4}$ is the van der Waals length scale and $E_{\rm vdW}=\frac{\hbar^2}{m\ell_{\rm vdW}^2}$ indicates the van der Waals energy scale.}
\label{fig1}
\end{figure}

\begin{figure}[h!]
\centering
 \includegraphics[scale=0.53]{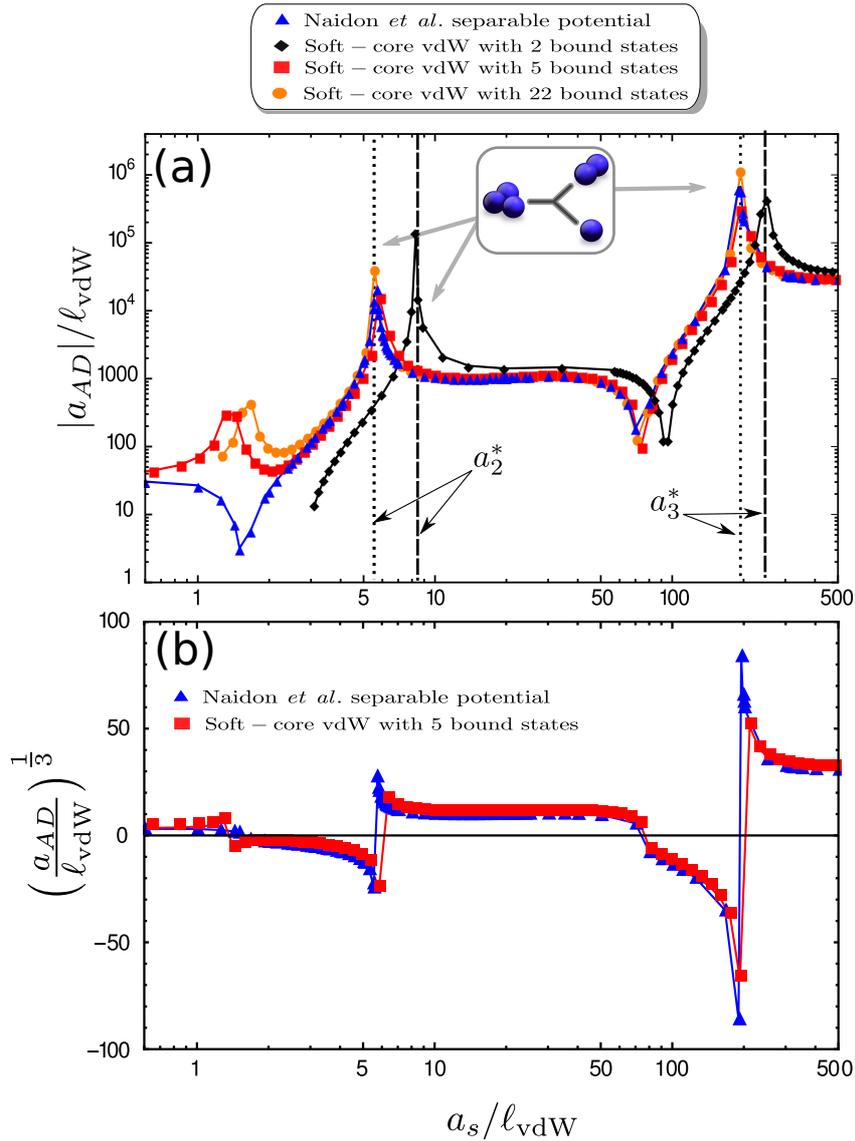}
 \caption{(color online) (a) The absolute value atom-dimer scattering length, $|a_{\rm AD}|/\ell_{\rm vdW}$, as a function of the two-body $s$-wave scattering length for different types of two-body potentials. The blue triangles correspond to Naidon {\it et al.} separable potential. The black diamonds, red squares and orange circles refer to the calculations for a soft-core vdW potential which contains in total 2, 5, and 22 bound states, respectively. 
 The vertical dotted, dashed and dashed-dotted lines indicate the positions of the atom-dimer resonances. 
 (b) An illustration of the sign dependence of the atom-dimer scattering length, $|a_{\rm AD}|/\ell_{\rm vdW}$ for the Naidon {\it et al.} separable approach (blue triangles) and for the soft-core vdW potential which contains $5$ bound states (red squares).
 Note that $\ell_{\rm vdW}=\frac{1}{2}(m C_6/\hbar^2)^{1/4}$ is the van der Waals length scale.}
\label{fig2}
\end{figure}

Fig.\ref{fig1} illustrates the dimer energies which are compared for different separable potentials, in order to highlight their regimes of validity.
In particular, the dimer energies which are computed within the separable potential based on the Naidon {\it et al.} approach are indicated by the blue solid line in Fig.\ref{fig1}.
Moreover, the red dotted and orange solid lines denote the dimer energies for a soft-core vdW potential which contains a total number number of 22 and 10 bound states, respectively.
Note, that the depicted dimer energies (red dotted and orange solid line) correspond to the least bound state which can be supported by the soft-core vdW potential.
The black dashed line denotes the universal dimer binding energies, i.e. $E=-\hbar^2/(m a_s^2)$.
For $a_s>25 \ell_{\rm vdW}$ the two-body binding energies calculated within the separable approximation are in excellent agreement with the binding energies for the universal dimer indicating in this manner the region where the two-body collisions are considered to be universal, namely independent of effective range corrections.
For scattering lengths $a_s<25 \ell_{\rm vdW}$ the effective range corrections become important and thus the binding energies of the separable potentials deviate from those of the universal dimer.
In addition, as the $s$ wave scattering length becomes $a_s<2.77 \ell_{\rm vdW}$ the binding energies for the Naidon {\it et al.} separable potential differ from those of a soft-core vdW interaction.
This designates the region of validity of the Naidon {\it et al.} separable potential.
More specifically, for $a_s< \ell_{\rm vdW}$ the Naidon {\it et al.} approach yields dimer energies which become shallower as the $s$-wave scattering length vanishes.
The main reason is that the separable potential in the Naidon {\it et al.} approach is constructed by the zero-energy two-body wavefunction.
Thus, it is expected that at energies away from the zero-energy bound state the potential cannot be regarded as non-local.

On the other hand, we observe that the soft-core vdW potential does not suffer from this restriction since the corresponding binding energies are obtained by diagonalizing the Lippmann-Schwinger kernel according to the prescription given in the previous subsection.

Fig.\ref{fig2}(a) depicts the absolute value of the elastic atom-dimer scattering length, $|a_{\rm AD}|$ as function of the $s$-wave scattering length $a_s$ in van der Waals units.
In particular the blue triangles indicate the atom-dimer scattering length where the the Naidon {\it et al.} separable potential is used.
The black diamonds denote the $a_{\rm AD}$ scattering length for the separable potential given in Eq.~(\ref{eq20}) which is constructed by a soft-core vdW interaction which contains 2 bound states in total.
Similarly, the red squares and the orange circles refer to the atom-dimer scattering lengths which correspond to soft-core potentials which contain 5 and 22 bound states, respectively.

Fig.\ref{fig2}(a) in a compact manner provides insights on the universal behavior of the atom-dimer scattering length.
Namely, the dependence of the atom-dimer scattering length for soft-core vdW interactions and van der Waals potential tail within the separable potential approximation.
Furthermore, for values of $s$-wave scattering length larger than $2 \ell_{\rm vdW}$, $a_s>2 \ell_{\rm vdW}$, the atom-dimer scattering length possesses resonant features for both types of separable potentials.
For $a_s<2 \ell_{\rm vdW}$ the atom-dimer scattering length exhibits pronounced features in the case of soft-core potentials which contain 5 (red squares) and 22 (orange circles) bound states.
These particular features are not associated with a resonant process and arise due to numerical instabilities.

In Fig.\ref{fig2}(a) we observe that the resonant features of $a_{\rm AD}$ for soft-core vdW potentials with only 2 bound states (black diamonds) are shifted to higher values of $a_s$ and as the number of total bound states is increased, i.e. red squares and orange circles, the atom-dimer resonances approach the results based on the Naidon {\it et al.} separable approximation which corresponds to a van der Waals tail containing an infinity of bound states.
This behavior mainly arises by the van der Waals character of the two-body potential. 
Namely, from a WKB viewpoint, the classical allowed region of soft-core potentials, see Eq.~(\ref{eq15}) can be divided into two regimes: (i) $r<\sigma$ the region close to origin where the potential is constant and (ii) the region of $r>\sigma$ where the attractive vdW potential tail prevails, i.e. the vdW region.
Therefore, as the soft-core potential becomes deeper the vdW region extends towards the origin mimicking in return a potential that possesses a pure van der Waals tail.
Thus, for deep soft-core potentials the atom-dimer collisions are strongly governed by the vdW physics exhibiting universal characteristics.
More specifically, as Fig.\ref{fig2}(a) shows the atom-dimer scattering length for a soft-core potential with more than 5 bound states (red squares and orange circles ) are in good agreement with the $a_{\rm AD}$ calculations of a pure van der Waals potential tail (blue triangles).

\begin{figure}[h!]
\centering
 \includegraphics[scale=0.65]{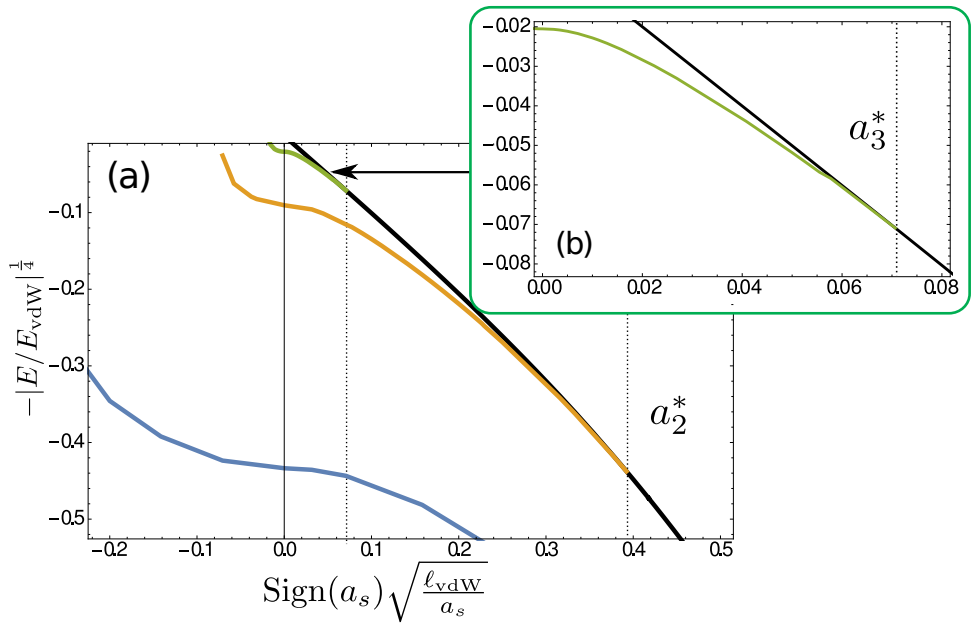}
 \caption{(color online) The biding energies, $E$, for the first three Efimov states as a function of the $s$-wave scattering length, $a_s$. (a) The ground Efimov state is indicated by the blue line. The $1^{\rm st}$ and $2^{\rm nd}$ excited states are denoted by the orange and green lines, respectively. The black solid line indicates the dimer threshold. (b) Zoom-in plot for the $2^{nd}$ excited Efimov state (green line) which crosses the dimer threshold (black line). The black dotted lines denote the position where the $1^{st}$  ($2^{nd}$) excited Efimov state crosses the dimer threshold, i.e. $a_s\equiv a_2^*$ ($a_s\equiv a_3^*$). Note that $\ell_{\rm vdW}=\frac{1}{2}(m C_6/\hbar^2)^{1/4}$ is the van der Waals length scale and $E_{\rm vdW}=\frac{\hbar^2}{m\ell_{\rm vdW}^2}$ indicates the van der Waals energy scale.}
\label{fig3}
\end{figure}

The underlying physical mechanism of the resonant features depicted in Fig.\ref{fig2}(a) at $a_s>2\ell_{\rm vdW}$ are associated with Efimov states.
Namely, as $a_s$ increases an Efimov state intersects with the dimer threshold; hence, the atom-dimer scattering length diverges at $a_{2}^*$ or $a_{3}^*$.
Panel(b) of Fig.\ref{fig2} demonstrates the sign change of the atom-dimer scattering length for the Naidon {\it et al.} separable potential (blue triangles) whereas the red squares indicate the corresponding calculation for the soft-core vdW potential with $5$ bound states.
In particular, Fig.\ref{fig2}(b) shows that for an increasing $a_s$ a trimer state is accessed, binding the atom and the dimer.
For $a_S<2~\ell_{\rm vdW}$ in the case of soft-core vdW potential with 5 bound states we observe that the corresponding enhancement possesses a positive $a_{AD}$ at small $a_s$ and as the $s$-wave scattering length increases $a_{AD}$ becomes negative.
Note that this behavior is not associated with a pole of the $a_{AD}$ but due to numerical instabilities.

In order to identify the Efimov states which are associated with the atom-dimer resonances of Fig.\ref{fig2}(a) in Fig.\ref{fig3} the first three Efimov states are illustrated.
More specifically, Fig.\ref{fig3} shows the Efimov trimer energies as a function of ${\rm Sign}(a_s)/\sqrt{a_s}$ in van der Waals units.
The bound spectrum is calculated by numerically solving the corresponding Faddeev equation employing the Naidon {\it et al.} separable potential \cite{naidonpra2014,naidonprl2014}.
The ground Efimov state, in Fig.\ref{fig3}(a), is indicated by the blue solid line, and the $1^{st}$ ($2^{nd}$) excited Efimov state is denoted by the orange (green) solid line whereas the black solid line refers to the dimer threshold.
Evidently, in Fig.\ref{fig3}(a)  we observe that, in contrast with the universal zero-range theory, the ground Efimov state does not cross the dimer threshold whereas the excited Efimov states [for the $2^{nd}$ excited state see Fig.\ref{fig3}(b)] merge with the dimer threshold at $a_{2}^*$ or $a_{3}^*$.
This suggests that the atom-dimer resonances in Fig.\ref{fig2}(a) are associated with excited Efimov states only within our model.

In addition, table~\ref{table:table1} contains information about the position of the atom-dimer resonances for both the soft-core and pure van der Waals potential tail which are depicted in Fig.\ref{fig2}(a) manifesting the universal behavior.
In particular for the position of the atom-dimer resonances which are associated with the $1^{st}$ excited Efimov state we observe that the results for soft-core vdW potentials with more than 5 bound states possess a difference less than $5.4\%$ from the corresponding calculations for a pure van der Waals tail, i.e the Naidon {\it et al.} separable potential.
Similarly, for the atom-dimer resonances which are associated with the $2^{nd}$ excited Efimov state the corresponding difference in the position of the resonances becomes less than $\sim 2\%$.
In addition, the third column of Table I indicates the ratio $\frac{a_3^*}{a_2^*}$ which differs from the corresponding ratio within the zero-range theory, namely $\frac{a_3^*}{a_2^*}=22.69$.
This difference between the zero-range theory and vdW separable potential approach is indicative of the effective range corrections which are absent in the zero-range potential approximation.
Furthermore, for the $a_2^*$ atom-dimer resonance the theoretical values are in reasonable agreement with the experimental ones.
Recall that in the $^{133}$Cs experiment of Ref.\cite{zenesinipra2014} two resonances are observed for two different Feshbach resonances (FR).
Namely for a narrow FR the position of the atom-dimer resonance is $a_2^*=(6.48\pm0.24)\ell_{\rm vdW}$ whereas the broad one the atom-dimer process is enhanced at $a_2^*=(4.15\pm0.09)\ell_{\rm vdW}$.

\begin{table}
 \caption{The position of the atom-dimer resonances in van der Waals for the potentials which are used in Fig.\ref{fig2}(a). The quantity $a_2^*$ ($a_3^*$) indicates the position of the atom-dimer resonance due to the $1^{st}$ ($2^{nd}$) excited Efimov state.}
\centering
    \begin{tabular}{llll} 
	\hline\noalign{\smallskip}
	~ &  $a^*_2$   & $a^*_3$& $ a^*_3/a^*_2$ \\[3pt]
	\tableheadseprule\noalign{\smallskip}
	\centering
	Soft-core with 2 bound states~& ~8.44~ & ~245.7&~29.11 \\ 
	Soft-core with 5 bound states	~&~5.930~ & ~197.6~&33.32\\ 
	Soft-core with 22 bound states~& ~5.609~ &~192.7~&~34.35\\ 
	The Naidon {\it et al.} separable potential	~&~5.628~ &~192.3~&~34.16 \\ 
	$^7$Li in Ref.\cite{machteyprl2012}~&~$6.03\pm0.12$~&~&~\\
	$^{133}$Cs in Ref.\cite{zenesinipra2014}~& $6.48\pm0.24$ \& $4.15\pm0.09$&~~&~\\
	\noalign{\smallskip}\hline
    \end{tabular}
    \label{table:table1}
\end{table}

\section{Summary}
In summary, the elastic atom-dimer collisions under the influence of two-body soft-core or pure van der Waals tail potential are studied.
The theoretical framework of our analysis is based on the integral Faddeev equations whereas the two-body interactions are modeled via suitable separable potentials.
The pure van der Waals potential tail is described using the Naidon {\it et al.} separable interaction where the zero-energy two-body wavefunction is utilized.
On the other hand, the factorization of the soft-core vdW potentials is based on the properties of the Hilbert-Schmidt expansion.
In particular, the form factor of the soft-core vdW potentials is obtained by calculating the two-body energy of the least bound state of the soft-core vdW interaction and its corresponding wavefunction.

Using these two types of separable potentials the universal aspects of the elastic atom-dimer collisions is then studied.
It is shown that the atom-dimer scattering lengths for soft-core vdW potentials with more than 2 bound states are in good agreement with the corresponding calculations for an infinitely deep vdW potential exhibiting the same resonant structure.
Furthermore, in order to identify the origin of the resonant features in the atom-dimer scattering length the Faddeev equation for a three-body bound system is solved.
Our analysis shows that the ground Efimov state does not cross the dimer threshold thus the resonant features in atom-dimer collisions are solely associated with the excited Efimov states.
In addition, highlighting the universal behavior of the atom-dimer elastic collisions we observe that: in the case of the $1^{st}$ excited Efimov states, the position of the resonances for soft-core vdW potential with more than 2 bound states differs by less than $5.4\%$ from the corresponding calculations with the Naidon {\it et al.} separable potential.
This difference decreases in the case of the $2^{nd}$ excited Efimov states where it becomes less than $2\%$.
Complementing our study similar conclusions were shown in Ref.\cite{li2016pra} for a pure vdW potential which contains only a finite number of bound states.
In addition, in Ref.\cite{li2016pra} shows that the Naidon {\it et al.} separable potential yields a three-body recombination rate for the ground Efimov state which does not follow the $a^4$ scaling law.
Beyond the simple model calculations presented here and in Ref.\cite{li2016pra}, Mestrom {\it et al.} in Ref.\cite{mestrom2016efimov} illustrates that the first excited Efimov states are strongly influenced by the higher partial wave interactions nearby the atom-dimer threshold.
In particular, Mestrom {\it et al.} demonstrate that an atom-dimer system colliding in the presence of $s$- and $d$-wave interactions force the ground and first excited Efimov states to not merge with the atom-dimer continuum. 

\begin{acknowledgements}
 We thank Jose D'Incao and Paul Mestrom for stimulating discussions related to the present investigation. This work was supported in part by NSF grant PHY-1607180.
\end{acknowledgements}
\bibliographystyle{unsrtnat}
\bibliography{atmdmbib_nn}

\end{document}